\documentclass{ifacconf}
\usepackage{amsmath}
\usepackage{graphicx}      % include this line if your document contains figures
\usepackage{natbib}        % required for bibliography
\usepackage{algorithm}
\usepackage{algorithmic}
\usepackage{amsfonts}
\usepackage{dsfont}
\usepackage{textcomp}
\usepackage{xcolor}
\usepackage{soul}

\newcommand{\pd}[2]{\frac{\partial #1}{\partial #2}}

\begin{document}
\begin{frontmatter}

\title{Variance Reduction of Quadcopter Trajectory Tracking in Turbulent Wind} 

\thanks[footnoteinfo]{This material is based upon work supported by the National Science Foundation under Grant No. 1925147. Any opinions, findings, and conclusions or recommendations expressed in this material are those of the author(s) and do not necessarily reflect the views of the National Science Foundation. Some of the computing for this project was performed at the High-Performance Computing Center (HPCC) at Oklahoma State University supported in part through the National Science Foundation grant OAC-1531128. We would like to acknowledge high-performance computing support from Cheyenne~\cite{computational2017cheyenne} (doi:10.5065/D6RX99HX) provided by NCAR's Computational and Information Systems Laboratory, sponsored by the National Science Foundation.}

\author[First]{Asma Tabassum} \author[First]{Rohit K. S. S. Vuppala}
\author[First]{He Bai}
\author[First]{Kursat Kara}
\address[First]{Oklahoma State University, Stillwater, OK 74078 USA \\E-mails: \{asma.tabassum,~rvuppal,~he.bai,~kursat.kara\}@okstate.edu.}

\begin{abstract}  % Abstract of not more than 250 words.
We consider a quadcopter operating in a turbulent windy environment. The turbulent environment may be imposed on a quadcopter by structures, landscapes, terrains and most importantly by the unique physical phenomena in the lower atmosphere. Turbulence can negatively impact quadcopter's performance and operations. Modeling turbulence as a stochastic random input, we investigate control designs that can reduce the turbulence effects on the quadcopter's motion. In particular, we design a minimum cost variance (MCV) controller aiming to minimize the cost in terms of its weighted sum of mean and variance. We linearize the quadcopter dynamics and examine the MCV controller derived from a set of coupled algebraic Riccati equations (CARE) with full-state feedback. Our preliminary simulation results show reduction in variance and in mean trajectory tracking error compared to a traditional linear quadratic regulator (LQR).

\end{abstract}


\end{frontmatter}
%===============================================================================

\section{Introduction}
Small Unmanned Aircraft System (sUAS) has become ubiquitous in diverse applications and are aggressively being integrated into the national airspace system (NAS). Multi-rotor platforms such as quadcopters have demonstrated significant potentials in small package delivery, surveillance operations and in many other applications. Many of the tasks involve operations in the low-altitude airspace. In the urban setting low-altitude operations impose challenges to operational and navigational tasks with its unique physical phenomena. Being under-actuated, a quadcopter is vulnerable to strong mean  wind velocity as well as unsteady wind gusts.~\cite{gill2017propeller} show that with a relative wind velocity more than $4-7 ms^{-1}$, the hover model of a quadcopter deteriorates. 

To  compensate for the wind effects, several disturbance rejection algorithms have been studied. Some of them require wind information, onboard wind estimation or prediction while others solve optimal policy without any wind information.~\cite{tran2015quadrotor} illustrate the performance of the traditional PID and LQR controllers for disturbance rejection where an offline computed look-up table is used to estimate wind components in the simulation.~\cite{wang2016trajectory} propose a hierarchical nonlinear control scheme for a quadcopter to track a 3D trajectory
subject to wind gust disturbances from a von Karman model. In~\cite{zhang2016three}, a three-dimensional
fuzzy PID control method for stabilizing attitude control and precise trajectory tracking control is implemented with wind gusts generated from a Dryden model in the simulation.~\cite{yang2017active} investigate attitude control via a dual closed-loop control framework where gust wind is considered  dynamic disturbances and estimated by an extended state observer.~\cite{ding2018robust} propose a linear active disturbance rejection control (LADRC) for stability control of a quadcopter under wind gusts with a linear extended state observer (LESO) as a compensator. A geometric adaptive controller is proposed in~\cite{bisheban2018geometric} and a numerical example is illustrated. An adaptive mass estimator and an adaptive neural disturbance estimator are derived in~\cite{sierra2019wind} that complement the action of a set of PID controllers stabilizing a sUAS under wind and variable payload. A second order sliding mode controller based on the super twisting algorithm (STA) with an observer is employed in~\cite{hamadi2019observer} to reject  wind perturbation. A real-time simulation study in wind is provided in~\cite{davoudi2020quad}.~\cite{tran2021adaptive} introduce Particle
Swarm Optimisation (PSO) based Adaptive Strictly Negative Imaginary (SNI) controller for unknown wind disturbance rejection. 

Almost every controller in the literature developed for wind disturbance rejection is focused on reducing the mean of the tracking error. In this paper, our objective is to incorporate stochastic properties of wind into a controller and reduce the variance of tracking error, which, to the best of the authors' knowledge, has not been considered in previous research.  In particular, we introduce a Minimum Cost Variance controller (\cite{sain1965minimal}) which is a special case of risk sensitive control (\cite{sain1995cumulants}) in the  quadcopter control paradigm. We consider a  quadcopter  model including a body drag effect and a stochastic differential dynamic model to assimilate the stochastic property of the wind. Previous studies except~\cite{davoudi2020quad} have not considered a realistic lower atmospheric conditions.~\cite{davoudi2020quad} is mostly focused on the realistic flight simulation in the wind field. Here, we adopt Large-Eddy Simulations to obtain high-fidelity Atmospheric Boundary Layer wind solutions and extract stochastic information. In the MCV formulation, the standard deviation of the wind information is incorporated into the stochastic model and an optimal controller is obtained to optimize the weighted sum of the mean and the variance of the cost. To generate a MCV controller for quadcopter trajectory tracking, we linearize the quadcopter dynamics along a planned trajectory and create a finite-horizon MCV controller based on the linearized model. We simulate hover, straight line and circular trajectories with the LES wind data to examine the effectiveness of the controller. In each case, we compare the MCV controller with an LQR controller and find that the MCV controller produces reduced turbulent effects and tracking error. %reduces turbulent and noise effect and decreases tracking error.

The rest of the paper is structured as follows. Section ~\ref{math} discusses the mathematical models of the wind and the quadcopter dynamics used for controller designs. In Section~\ref{mcv} we present our controller design.  We discuss the simulation results in Section~\ref{result}. Future work is summarized in Section~\ref{future}.

\section{Mathematical Model} \label{math}
\subsection{Modeling Atmospheric Wind Effects}
For control designs, we model a wind velocity in the inertial frame, $v_{w} \in {\rm I\!R}^3$, as the summation of a mean component ($\bar v_w $) and a  stochastic component ($ \tilde v_{w}$)
\begin{equation}
    v_w = \bar v_w + \tilde v_{w}.\label{eq:wind}
\end{equation}
Stochastic formulations of $\tilde v_{w}$ like~\cite{von1948progress} and its variants are majorly dependent on canonical spectral energy function for incorporating disturbances or gusts in the wind field. To simplify the formulation, we model $\tilde v_w$ as a zero-mean Gaussian distribution noise. For quadcopter operations with limited range and duration, the wind is assumed to be spatially-temporally homogeneous, which means that $\bar v_w$ and the statistics of $\tilde v_w$ are independent of time and location.

We note that in reality, Atmospheric Boundary Layers are characterised by more complex highly coherent turbulent structures. Hence, using stochastic models might lead to significant differences between realistic and predicted wind field conditions. Therefore, we adopt Large-Eddy Simulations (LES) in our simulations to generate high-fidelity Atmospheric Boundary Layer solutions that accurately capture the unsteady highly coherent eddies at various scales, important for closely depicting realistic wind field. Such LES wind data are used in simulations to validate our  controllers that assume a Gaussian distribution on the turbulence $\tilde v_w$. Details of the LES data can be found in Section~\ref{sec:LES}.

\subsection{Quadcopter Dynamic Model under Wind Disturbance } \label{Sec:quad model}
We consider a quadcopter aerial vehicle as a single rigid-body with four identical rotors. Let $p \in {\rm I\!R}^{3}$ be its inertial position, $q=[q_w,q_x,q_y,q_z]^T \in {\rm I\!R}^{4}$ the unit quaternion representing its orientation in the inertial frame, and $v \in {\rm I\!R}^{3}$ the inertial velocity. Considering the quadrotor under wind disturbance, the system dynamics for the quadcopter is given by
\begin{align}
   \label{eqn:position} \Dot{p}&= v + v_{w} \\ 
    \Dot{q}&= \frac{1}{2}q \otimes \begin{bmatrix} 
    0 \\ \label{eqn:angle}
    \omega
    \end{bmatrix} \\
    \Dot{v} &= \boldsymbol{g} + \frac{1}{m}q \odot \boldsymbol{f_{c}}  -  \frac{1}{m} f_{D} \label{eqn:velocity}
\end{align}
where $\odot$ and $\otimes$ are the quaternion rotation and multiplication, respectively, $v_{w} \in {\rm I\!R}^3$ is the wind velocity in the inertial frame as given in~\eqref{eq:wind}, %that comprises of a mean component ($\bar v_w $) along with a  stochastic component ($ \tilde v_{w}$),
$\boldsymbol{g}=[0;0;-g]^{T}$ represents the gravitational acceleration, $ f_{D} \in {\rm I\!R}^3$ is the drag force on the quadcopter in the inertial frame and $m$ is the mass of the quadcopter. Here, $\omega \in {\rm I\!R}^3$ is the  angular rate represented in the body frame and $\boldsymbol{f_{c}} \in {\rm I\!R}^3$ is the collective thrust in the body frame  given by 
\begin{equation*}
    \boldsymbol{f_{c}} = \begin{bmatrix}
    0 \\
    0\\
    f_{c}
    \end{bmatrix}.
\end{equation*}
Let $u=[\omega^T,f_c]$, which is considered the system input for control design. Once $u$ is designed, a low-level controller for rotor speed control can be used to track $u$. 

We assume that the drag force $f_D$ in the inertial frame is of the following form 
\begin{equation}
   f_{D}= RDv_B \lvert \lvert v_B \rvert \rvert =RDR^Tv\lvert \lvert R^Tv \rvert \rvert= \lvert \lvert v \rvert \rvert RDR^Tv
\end{equation} 
where $R\in SO(3)$ is the orientation matrix represented by $q$, $v_{B}=R^Tv$ is the relative air velocity in the body frame and $D$ is the drag coefficient matrix expressed as
\begin{equation*}
    D= \begin{bmatrix}
    d_{x} & 0 & 0 \\
    0 & d_{y} & 0 \\
    0 & 0 & d_{z}
    \end{bmatrix}.
\end{equation*}
This drag model is adapted from a standard 1D drag model $f_D=dv^2$ for some constant $d$. The orientation matrix $R$ is calculated from $q$ using
\begin{equation}
    \Bar{Q}^{\times T}(q) Q^{\times}(q) = \begin{bmatrix}
    1 & \boldsymbol{0} \\
    \boldsymbol{0} & R
    \end{bmatrix} 
\end{equation}
where
\begin{equation*}
  Q^{{\times}}(q) =  \begin{bmatrix}
    q_{w} & -q_{x} & -q_{y} & -q_{z} \\
    q_{x} & q_{w} & -q_{z} & q_{y} \\
    q_{y} & q_{z} & q_{w} & -q_{x} \\
    q_{z} & -q_{y} & q_{x} & q_{w} 
     \end{bmatrix} 
\end{equation*}
and
\begin{equation*}
  \Bar{Q}^{{\times}} (q)=  \begin{bmatrix}
    q_{w} & -q_{x} & -q_{y} & -q_{z} \\
    q_{x} & q_{w} & q_{z} & -q_{y} \\
    q_{y} & -q_{z} & q_{w} & q_{x} \\
    q_{z} & q_{y} & -q_{x} & q_{w} 
     \end{bmatrix}. 
\end{equation*}

\section{Minimum Cost Variance Controller} \label{mcv}
\subsection{Review of minimum cost variance control}
Even though LQR controllers have been proven to be a good choice for tracking problems, unfortunately the solution derived is independent of noise statistics. The optimal solution is deduced considering the mean of the quadratic cost while ignoring the higher order information and is indifferent to stochasticity according to the uncertainty equivalence principle. The necessity of considering the higher order momenta is to address robustness and reduce fluctuation in the trajectory due to stochastic disturbances.~\cite{sain1965minimal} introduces a minimum cost variance  controller, a special case of cost cumulant control that minimizes a given cost in terms of its mean and variance at a level decided by the user or performance requirement.  Preliminary investigations on MCV and its connection to cost cumulant control and traditional Linear Quadratic Gaussian (LQG) controller are discussed in~\cite{sain1995cumulants}. Coupled algebraic Riccati equations has been solved for full-state feedback MCV and sufficient conditions for the existence and uniqueness of solutions for finite horizon and infinite horizon were established in \cite{sain1995cumulants} and \cite{won2003infinite}, respectively.

Consider a generic linear stochastic dynamic system with state $x\in {\rm I\!R}^{n}$ and input $u\in {\rm I\!R}^{m}$ given by
\begin{equation} \label{eqn:LTI}
    dx = (Ax+ Bu)dt + Gdw.
\end{equation}
The system matrices $A \in {\rm I\!R}^{n\times n}$, $B \in {\rm I\!R}^{n \times m}$ and $G \in {\rm I\!R}^{n\times s}$ are known, where $n$, $m$ and $s$ are the number of state, input and noise, respectively. The stochastic noise $dw$ represents a stationary Wiener process and satisfies
\begin{equation}\label{eq:W}
    E[(w(t_{1})-w(t_{2}))(w(t_{1})-w(t_{2}))^{T}]=W|t_{1}-t_{2}|
\end{equation}
where $E[\cdot]$ denotes the expectation function and $W\in {\rm I\!R}^{s \times s}$ is a positive definite matrix. A traditional quadratic cost function has a form
\begin{equation}
    J_{\infty}(x,u,t_{f})= \int_{0}^{t_{f}} (x^{T}Qx+ u^{T}Ru) d\tau \label{cost function generic}
\end{equation}
The objective of the MCV controller is to find optimal policy such that it minimizes the weighted sum of mean and variance of the cost function given by~\eqref{cost function generic}. Hence the objective function is as follows:
\begin{equation}
    \textit{j}_{\infty}(x,u) = \lim_{t_{f} \rightarrow{\infty}} \frac{E[J_{\infty}(x,u,t_{f})] }{t_{f}} + \gamma \lim_{t_{f} \rightarrow{\infty}} \frac{Var[J_{\infty}(x,u,t_{f})] }{t_{f}} \label{objective infinite}
\end{equation}
where $Var[\cdot]$ denotes the variance and $\gamma$ is a positive parameter that regulates the variance in the objective minimization. The higher the value of $\gamma$, the smaller the variance component in the optimal solution. Equation~\ref{cost function generic} and~\ref{objective infinite} are for the infinite horizon formulation. 

For a finite horizon optimal control problem, we consider the following stochastic differential equation, 
\begin{equation}
    dx(t)=  (A(t)x(t)+B(t) u(t))dt + G(t) dw(t) \label{eqn:time SDE}
\end{equation}
where $A(t)$  and $B(t)$ are the linearized state matrices about the nominal trajectory at time $t$, $dw(t)$ represents a stationary Wiener process same as \eqref{eq:W} and $G(t) \in {\rm I\!R}^{n \times s}$ . The cost and the objective equations are modified as
\begin{equation}
        J(x,u,t_{f})= \int_{0}^{t_{f}} (x^{T}(t)Q(t)x(t)+ u^{T}(t)R(t)u(t)) d\tau + Q_{f} \label{cost function finite}
\end{equation}
where $Q_{f}$ is the terminal cost and
\begin{equation}
    \textit{j}(x,u) =  E[J(x,u,t_{f})] + \gamma Var[J(x,u,t_{f})]. \label{objective finite}
\end{equation}

We utilize the following  two lemmas to solve for the infinite and finite horizon optimal controllers, respectively.
\begin{lem}~(\cite{won2003infinite})
The optimal control gain for the infinite horizon optimal control  problem \eqref{eqn:LTI}--\eqref{objective infinite} has the form
\begin{equation}
     K= - R^{-1}B^{T}(M+\gamma H) \label{gain: MCV}
\end{equation} where $\gamma > 0$ and $M$ and $H$ satisfy the following CARE:
\begin{equation}
    A^{T}M+MA+Q-MBR^{-1}B^{T}M+\gamma^{2}HBR^{-1}B^{T}H=0 \label{care}
\end{equation}
\begin{multline}
 A^{T}H+HA-MBR^{-1}B^{T}H- HBR^{-1}B^{T}M \\ - 2\gamma HBR^{-1}B^{T}H+4MGWG^{T}M=0.   
\end{multline}
\label{lemma infinite}
\end{lem}

\begin{lem}~(\cite{sain1965minimal})
The optimal control gain for the finite horizon control problem \eqref{eqn:time SDE}--\eqref{objective finite} has the form
\begin{equation}
 K(t)=-R^{-1}(t)B^{T}(t)(M(t)+\gamma H(t)) \label{gain:MCV time}
\end{equation} 
where $\gamma > 0$ and $M(t)$ and $H(t)$ satisfy
\begin{multline}
 \dot  M(t)+ A^{T}(t)M(t)+M(t)A(t)\\+Q(t)-M(t)B(t)R(t)^{-1}B(t)^{T}M(t) \\ + \gamma^{2} H(t) B(t)R^{-1}(t) B(t)^{T}H(t)=0   \label{time care 1}
\end{multline}
\begin{multline}
 \dot H(t)+   A(t)^{T}H(t)+H(t)A(t)+4M(t)G(t)W(t)G^{T}(t)M(t) \\-M(t)B(t)R^{-1}(t)B(t)^{T}H(t)
 \\-H(t)B(t)R^{-1}(t)B(t)^{T}M(t)\\-2\gamma H(t)B(t)R^{-1}(t)B^{T}(t)H(t)=0  \label{time care 2}
\end{multline}
with boundary condition $M(t_{f})=Q_{f}$ and $H(t_{f})= 0$. \label{lemma finite}
\end{lem}

Once the feedback gain matrix $K$ ($K(t)$) is found in Lemma~\ref{lemma infinite} (Lemma~\ref{lemma finite}), the controller of the form $u=u_{n}+ K(x-x_{n})$ ($u(t)=u_{n}(t)+ K(t)(x(t)-x_{n}(t))$) is implemented in~\eqref{eqn:LTI}, where $x_n$ and $u_n$ are the reference state and input, respectively.

%$u_{n}$ is the reference input. 

\subsection{Application to sUAS control}
Let $x=[p^T,q^T,v^T]^T$. To create a MCV controller we linearize the quadrotor dynamics~\eqref{eqn:position}--\eqref{eqn:velocity} in Section~\ref{Sec:quad model} to obtain a linearized system as in~\eqref{eqn:LTI} and \eqref{eqn:time SDE}. The linearized $A$ and $B$ matrices are given by
\begin{equation}
    A  =  \begin{bmatrix}
    \boldsymbol{0} & \boldsymbol{0} & \frac{\partial}{\partial v} \dot p \\
    \boldsymbol{0}&\frac{\partial}{\partial q} \dot{q}& \boldsymbol{0}\\
    \boldsymbol{0}&\frac{\partial}{\partial q} \dot{v} & \frac{\partial}{\partial v} \dot{v}
    \end{bmatrix} \label{Amatrix}
\end{equation}
\begin{equation}
    B = \begin{bmatrix}
    \boldsymbol{0} & \boldsymbol{0}~ \\
    \frac{\partial}{\partial \omega} \dot{q}& \boldsymbol{0} \\
    \boldsymbol{0} &\frac{\partial}{\partial f_{c}} \dot{v} 
    \end{bmatrix}\label{Bmatrix}
\end{equation} 
where $\boldsymbol{0}$ implies that the partial derivative of the associated matrix entries are zero. Because a unit quaternion induces a constraint on the respective states so that $\lvert \lvert q \rvert \rvert = 1$, we make use of a  special quaternion $q_{u}= q \cdot \lvert \lvert q \rvert \rvert^{-1}$ as described in~\cite{foehn2018onboard} and derive the partial derivatives as

\begin{equation}
   \frac{\partial f(q_{u})}{\partial q}= \frac{\partial f(q)}{\partial q_{u}}\cdot \frac{\partial}{\partial q} (q \cdot \lvert \lvert q \rvert\rvert^{-1})
\end{equation}where 
\begin{equation}
   \frac{\partial}{\partial q} (q\cdot \lvert \lvert 
   q \rvert \rvert^{-1}) =  (I_{4}- \lvert \lvert q \rvert \rvert^{-2} qq^{T}) \lvert \lvert q\rvert \rvert^{-1}.
\end{equation}

The partial derivatives of \eqref{Amatrix} and \eqref{Bmatrix} are
\begin{equation}
\frac{\partial}{\partial v}\dot{p} = I_{3}
\end{equation}
\begin{equation}
    \frac{\partial}{\partial q} \dot{q}= \frac{1}{2} \begin{bmatrix}
    0 & -\omega_{x} & -\omega_{y} & - \omega_{z}\\
\omega_{x}& 0 & \omega_{z}&  -\omega_{y} \\ 
\omega_{y}& -\omega_{z}& 0&\omega_{x} \\
\omega_{z} & \omega_{y}& -\omega_{x} & 0
\end{bmatrix} (I_{4}- \lvert \lvert q \rvert \rvert^{-2} qq^{T}) \lvert \lvert q\rvert \rvert^{-1}
\end{equation}
\begin{equation}
    \frac{\partial}{\partial q}\dot{v} = 2f_{c} \begin{bmatrix}
    q_{y} & q_{z} & q_{w} & q_{x} \\
    -q_{z} & -q_{w} & q_{z} & q_{y} \\
    q_{w}& -q_{x} &-q_{y} &q_{z} \\
    \end{bmatrix} (I_{4}- \lvert \lvert q \rvert \rvert^{-2} qq^{T}) \lvert \lvert q\rvert \rvert^{-1}
\end{equation}

\begin{equation}
\frac{\partial}{\partial v}\dot{v} = RDR^T(\lvert \lvert v \rvert \rvert I_3 +  \frac{vv^T}{\lvert \lvert v\rvert \rvert})
\end{equation}

\begin{equation}
    \frac{\partial}{\partial \omega}\dot{q}= \frac{1}{2} \begin{bmatrix}
    -q_{x} & -q_{y} & -q_{z} \\
    q_{w} & -q_{z} & -q_{y} \\
    q_{z} & q_{w} & q_{x} \\
    -q_{y} & q_{x} & q_{w}
    \end{bmatrix}
\end{equation}
\begin{equation}
    \frac{\partial}{\partial f_{c}}\dot{v}= \begin{bmatrix}
    q_{w}q_{y}+q_{x}q_{z} \\
    q_{y}q_{z} - q_{w}q_{x} \\
    q_{w}^{2}-q_{x}^{2} - q_{y}^{2}+ q_{z}^{2}
    \end{bmatrix}.
\end{equation}

In this work, the 3D turbulent wind $\tilde v_{w}$ is considered the stochastic noise. Therefore, we obtain  $G \in {\rm I\!R}^{10 \times 3}$  in the linearized dynamics~\eqref{eqn:LTI} from~\eqref{eq:wind} and~\eqref{eqn:position} as
\begin{equation}
    G =\begin{bmatrix}
    I_3\\
    \mathbf{0}
    \end{bmatrix}.
\end{equation}
The $W$ matrix in~\eqref{eq:W} is chosen to be the covariance matrix of $\tilde v_w$. Note that the choice of $G$ and $W$ is not unique. We may also choose $W=I_3$ and set the first three diagonal elements in $G$ as the standard deviation of the wind in each direction. The mean wind $\bar v_w$ is considered a deterministic disturbance to the linearized system. 

%For the time varying problem, we let $G(t)=G$ and $W(t)=W$. 

The linearized system is evaluated at the corresponding reference  trajectory and control ($x_{n},u_{n}$). In particular, the reference for the quaternion $q$ and the velocity $v$ is $[1,0,0,0]^T$ and $-\bar v_{w}+\dot p_n$, respectively, where $p_n$ is the reference trajectory for the state $p$. For $p_n$, we consider two scenarios. For hovering control, we use the infinite horizon formulation and choose $p_n$ as the hovering point. The resulting time-invariant linear system is described  by $(A,B,G,W)$. For trajectory tracking, we use the finite horizon formulation and choose $p_n$ as the nominal trajectory. The resulting time-varying linear system is described  by $(A(t),B(t),G(t),W(t))$, where $G(t)=G$ and $W(t)=W$. 

We generate the reference control $u_{n}$ by finding a stable gain $K$ (through LQR or MCV) at the first linearization point and then setting $u_{n}= u^{0}+ K(x-x_{n})$, where $u^0 = [0~0~0~mg]^T$. %considering a hovering scenario with initial input $u_{0}$ about the first reference point $x_{1}$, linearize the system about $(x_{1},u_{0})$ and calculate reference stable input such that  

Note that the information of $\bar v_w$ is used for linearization while the statistics of $\tilde v_w$ is used in $G$ (or $W$). The information of $\bar v_w$ and statistics of $\tilde v_w$ may be provided by measurements from available wind towers or  wind estimation algorithms onboard the quadcopter.

 In the infinite horizon problem (for hovering control), the solution for $M$ and $H$ in~\eqref{gain: MCV} can be obtained by iteratively solving
\begin{multline}
      (A+BK_{k})^{T}M_{k}+ M_{k}(A+BK_{k}) \\ + K_{k}^{T}RK_{k}+Q= 0 \label{forward:M}
\end{multline}
\begin{multline}
       (A+BK_{k})^{T}H_{k}+ H_{k}(A+BK_{k})\\+ 4M_{k}GWG^TM_{k}=0. \label{forward:H}
\end{multline}
Algorithm \ref{algorithm infinite}, given in ~\cite{won2003infinite},  is used to find the optimal policy.

\begin{algorithm}
\caption{Iterative Infinite Horizon MCV Control}
\label{algorithm infinite}
\begin{enumerate}
    \item \textbf{Given}  Linearized dynamics $A$, $B$, $G$, $W$, $\gamma$, cost terms $Q,R$ and a threshold $\epsilon>0$
    \item \textbf{Initialization} Let $k=0$ and choose initial stable gain $K_{0}$
    \item Obtain $M_{k}$ and $H_{k}$ by solving \eqref{forward:M} - \eqref{forward:H}.
    \item Compute $K_{k+1}$ from \eqref{gain: MCV}
    \item Evaluate 
    \begin{equation}
        \sigma:= \frac{\lvert \lvert K_{k+1} - K_{k}\rvert \rvert}{\lvert \lvert K_{k} \rvert \rvert}
\end{equation}
    \item \textbf{if} $\sigma > \epsilon $, $k$ \textleftarrow $k+1$ and go back to step 3\\
    \textbf{else} optimal gain found
\end{enumerate}
\end{algorithm}

In the finite horizon control (for trajectory tracking), the solution involves solving \eqref{time care 1}--\eqref{time care 2} backward in time and then calculating the time-varying gain $K(t)$ in~\eqref{gain:MCV time} forward in time. The algorithm is presented in Algorithm~\ref{algorithm finite}.

\begin{algorithm}
\caption{Finite Horizon MCV Control}
\label{algorithm finite}
\begin{enumerate}
        \item \textbf{Given} Linearized system matrices $A(t)$, $B(t)$, $G$, $W$ along the reference trajectory $x_{n}$ and reference control $u_{n}$, $\gamma$ and cost terms $Q$, $R$ and $Q_{f}$
       % \item \textbf{Initialization} Start with stable control $u_{n}$
    \item Set $M(t_f)$=$Q_{f}$ and $H(t_f)=0$ and solve time varying CARE equations \eqref{time care 1} and \eqref{time care 2} backward in time
  
    \item Calculate $K(t)$ from \eqref{gain:MCV time} forward in time.
\end{enumerate}
\end{algorithm}
We generate the two controllers for the nonlinear quadcopter dynamics in simulations and evaluated the performance in the next section.

\section{Simulations and Result Analysis}\label{result}
\subsection{Large-Eddy Simulation for Wind Field}\label{sec:LES}
\subsubsection{Governing Equations}
\hfill

For simplicity, dry adiabatic atmospheric conditions are considered for the idealized simulations. Hence, we only present the governing equations and methodology corresponding to these specific conditions. Cloud Model 1 (CM1) described in \cite{bryan2002benchmark} was employed for numerical simulation, integrating the governing equations for $u,v,w,\pi',\theta' $, where $\pi'$ is the non-dimensional pressure, $\theta'$ is the potential temperature deviations from the base state (represented by subscript ``0") which is in hydrostatic balance and $(u,v,w)$ represent the three-dimensional (3D) wind velocity field in the inertial frame. The ideal gas equation $p=\rho RT$ is used for the equation of state. The governing equations are:
\begin{align}
    \pd{u}{t} + c_p \theta_p \pd{\pi'}{x} 
    &= adv(u) + fv + T_u + N_u 
    \label{x-mom} \\
    \pd{v}{t} + c_p \theta_p \pd{\pi'}{y} 
    &= adv(v) - fu + T_v + N_v 
    \label{y-mom} \\
    \pd{w}{t} + c_p \theta_p \pd{\pi'}{z} 
    &= adv(w) + B + T_w  + N_w 
     \label{z-mom}
\end{align}
\begin{align}
    \pd{\theta'}{t} &= adv(\theta) + T_{\theta} + N_{\theta} + \dot{Q_{\theta}} \label{theta'}\\
    \pd{\pi'}{t} &= adv(\pi) - \frac{R}{c_v} \pi \left( \pd{u}{x} + \pd{v}{y} + \pd{w}{z}\right) + \dot{Q_{\theta}} \label{pi'}
\end{align}
where `adv()' represents the advection operator for a generic variable $\alpha$ given as
\begin{align*}
     adv(\alpha) = -u \pd{\alpha}{x} -v \pd{\alpha}{y} -w \pd{\alpha}{z},
\end{align*}
where, $T,\dot{Q_{\theta}}$ represent the tendencies from turbulence and external tendencies to internal energy (radiative cooling/heating). Furthermore, the terms $N$, $f$, and $B$ represent the Newtonian Relaxation parameter, Coriolis parameter and buoyancy, respectively. The turbulence tendencies in the equations could be expressed as (writing in the Einstein notations using $(i,j=1,2,3)$ and $(x_1 = x,x_2=y,x_3=z;u_1 = u,u_2=v,u_3=w )$), 
\begin{align}
    T_{u_{(i)}} = \frac{1}{\rho} \left[
    \pd{\tau_{ij}}{x_j}  \right],\quad
    T_{\theta} = - \frac{1}{\rho} \left[
    \pd{\tau^{\theta}_{i}}{x_i}  \right]. 
\end{align}
The subgrid-stress terms are formulated as below:
\begin{align}
    \tau_{ij} \equiv \overline{\rho u'_i u'_j} = 2\rho K_m S_{ij}\\
    \tau^{\theta}_{i} \equiv \overline{\rho u'_{i} \theta'} = -K_h\rho \pd{\theta}{x_i}
\end{align}
 where $S_{ij}$ is the strain tensor, $K_m$ is the viscosity, $K_h$ is the diffusivity, and $K_m,K_h$ are determined from the type of subgrid closure used like TKE (Turbulence Kinetic Energy) similar to \cite{deardorff1980stratocumulus} or Smagorinsky from \cite{smagorinsky1963general}.
\subsubsection{Numerical Simulation Setup} 
\hfill

\begin{figure}[h]
    \centering
    \includegraphics[width=7cm]{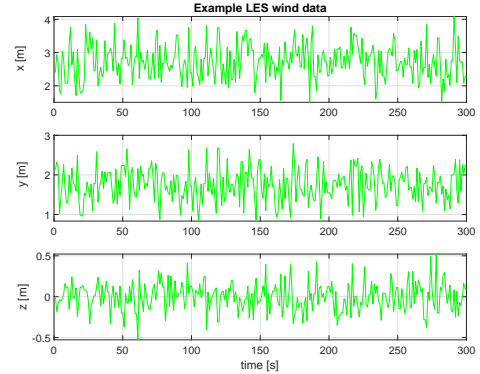}
    \caption{LES wind velocity in $x$, $y$ and $z$ direction. Wind in the $x$ direction has the highest mean and variance and the $z$ direction has the lowest mean and variance. This data is extracted at the position $(1,1,8)$ meters.}
    \label{fig:wind}
\end{figure}

The simulation was setup by closely following \cite{beare2006intercomparison}, for a stable or nocturnal boundary layer case. The computational domain has dimensions of $400$ m $\times$ $400$ m $\times$ $400$ m and isotropic grid resolution of 3.125 m in all the three directions. The geostrophic wind was set as 8 $ms^{-1}$ in the East-West direction with a Coriolis parameter of $1.39$ $\times$ $10^{-4}~s^{-1}$ (73$^\circ$ N). Surface cooling of 0.25 K $h^{-1}$ was employed and potential temperature profile was initialised with a mixed layer up to 100m with a value of 265K and overlying inversion strength of 0.01 K $m^{-1}$. Turbulent kinetic energy (TKE) closure for sub-grid scale terms was employed and the TKE was initialised as $0.4(1 - z/250)^3~ m^2 s^{-2}$ below a height of $250$ m, where $z$ represents the height. Periodic boundary conditions in the 4 sides, with no-slip at the bottom and slip at the top, were considered. The fifth order weighted essentially non-oscillatory (WENO) scheme with implicit diffusion from \cite{jiang1996efficient} was used. The wind data was collected from 8hr to 9hr in the simulation time after it reached to a quasi-equilibrium state. The wind was generated every 1 second. An example of wind velocity magnitude is shown in the Fig~\ref{fig:wind}. 

\subsection{Tracking with LES Wind}
%We solve for optimal control policy for trajectory tracking in addition to hover maintenance. 
 We incorporate LES turbulence wind  to test and validate the controller designs. For the drag coefficient $D$, we use~\cite{allison2020wind} 
\begin{equation}
    D= \min(1.1,(0.2+0.9 \exp{(-0.6\lvert\lvert v_{w}-\dot p \rvert \rvert-2)}))I_{3}
\end{equation} where $I_{3}$ is the identity matrix. For simplicity, we consider $m=1$ kg. For trajectory tracking, the initial conditions are $x_0=[0,~0,~0,~1,~0,~0,~0,~0.001,~0,~0]^{T}$ and $u_0=[0,~0,~0,~10]^{T}$. Note that we start with at least one non-zero entries of velocity $v$  so that we do not get division by zero error from drag component of \eqref{eqn:velocity} during linearization. Since we are only dealing with trajectories at the altitude lower than $8$ m, we extracted $10$ minutes of LES data around our nominal trajectory points. The mean wind velocity of the extracted wind data is $\bar v_{w}=[2.72,~1.752,~-0.006]^T$ m/s.

We compare our results with a traditional LQR architecture using the same dynamics described in Section~\ref{Sec:quad model}. We choose the cost such that the disturbance free trajectory matches the nominal trajectory closely. The quadratic cost for every simulation is fixed at $Q=diag([10,10,10,1,1,1,1,0.1,0.1,0.1])$ and $R=diag([1,5,5,0.1])$. We set the final cost is set at $Q_{f}=diag([20,20,20,0.1,0.1,0.1,0.1,0.1,0.1,0.1])$, for the finite horizon controller design.

\begin{figure}[h]
    \centering
    \includegraphics[width=7cm]{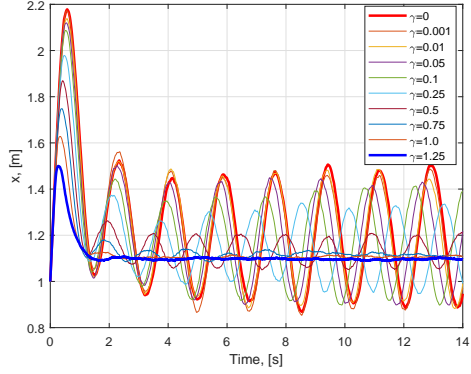}
    \caption{Change in trajectory in hover with increasing $\gamma$. Red indicates the LQR trajectory $(\gamma=0)$ and Blue indicates the MCV trajectory with  $(\gamma=1.25)$. Increasing $\gamma$ reduces the variance.}
    \label{fig:gamma vs xtraj}
\end{figure}

\begin{figure}[h]
    \centering
    \includegraphics[width=7cm]{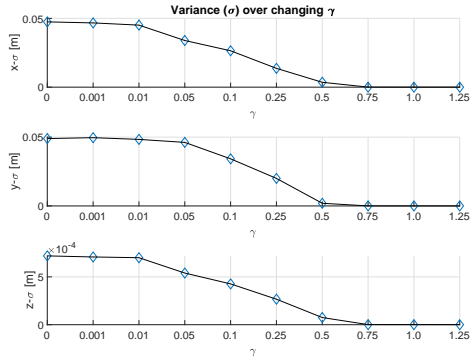}
    \caption{Change in trajectory variance in $x$, $y$, and $z$ directions with $\gamma$. As the $\gamma$ value increases, the variance decreases.}
    \label{fig:gamma vs var}
\end{figure}

\begin{figure}[h]
    \centering
    \includegraphics[width=7cm]{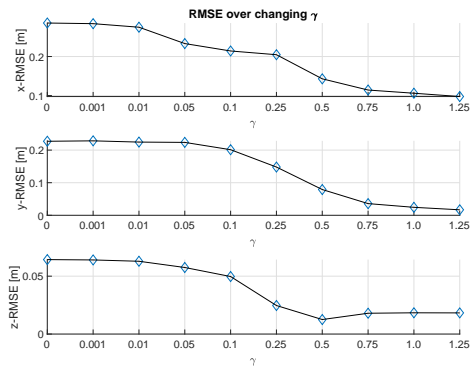}
    \caption{RMS error vs $\gamma$ illustrating the decrease in error with increasing $\gamma$.}
    \label{fig:gamma vs rms}
\end{figure}

To compare the effect of the tuning parameter $\gamma$, we simulate a  hovering scenario at the position $(1,1,8)$ meters with multiple $\gamma$. We show the evolution of $p_x$ trajectory with varying $\gamma$ in Fig~\ref{fig:gamma vs xtraj}, where we observe that the maximum value $\gamma=1.25$ significantly reduces the variances in the trajectory.  Figs~\ref{fig:gamma vs var} and~\ref{fig:gamma vs rms} show that increase in $\gamma$ value results in decrease in the variance and the Root Mean Square (RMS) error. Only in the $z$ direction, there is minuscule increase in the RMS error, however the values are still significantly lower comparing to the smaller $\gamma$ values. %Figure~\ref{fig:input} shows a comparison of the input signal $u$.


We next employ the finite horizon MCV controller illustrated in Algorithm~\ref{algorithm finite} for trajectory tracking problems. For the reference trajectories, we choose a straight line trajectory and a circular trajectory generated from minimum snap trajectories described in~\cite{5980409}. 

\begin{figure}[h]
    \centering
    \includegraphics[width=7cm]{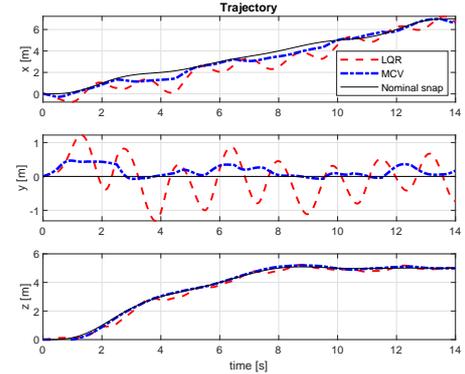}
    \caption{Comparison of straight line trajectory tracking with the LQR and the MCV controllers. Black is the nominal snap trajectory, blue corresponds to the MCV and red corresponds to the LQR trajectory. The deviation is smaller with the MCV controller.}
    \label{fig:traj_les}
\end{figure}

\begin{figure}
    \centering
    \includegraphics[width=7cm]{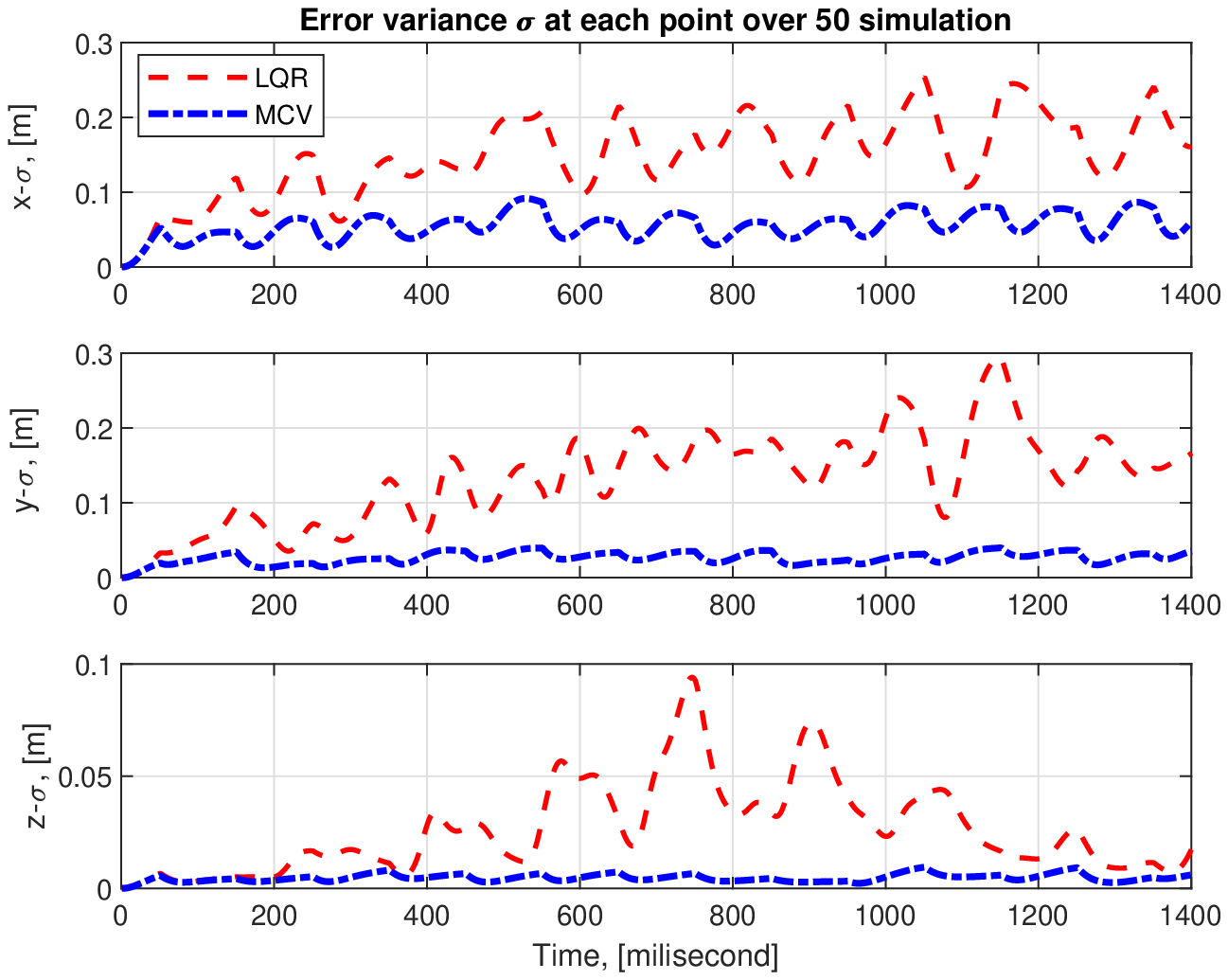}
    \caption{Comparison of error variance calculated at each point over 50 Monte Carlo simulations between the MCV and the LQR controllers for the straight line trajectory tracking.}
    \label{fig:varles}
\end{figure}

\begin{figure}
    \centering
    \includegraphics[width=7cm]{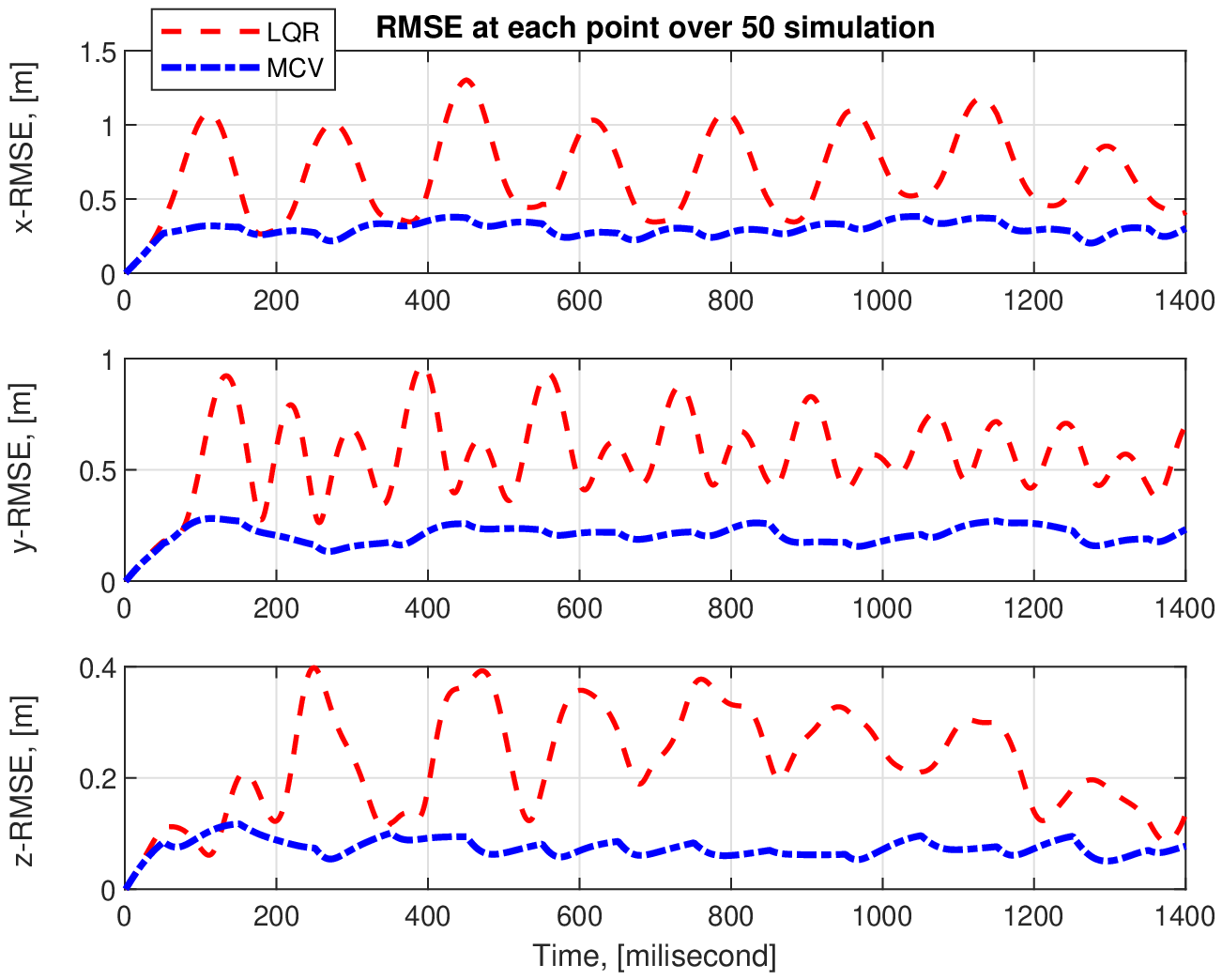}
    \caption{Comparison of RMSE calculated at each point over 50 Monte Carlo simulations  between the MCV and the LQR controllers for the straight line trajectory tracking.}
    \label{fig:rmsles}
\end{figure}

\begin{figure}
    \centering
    \includegraphics[width=7cm]{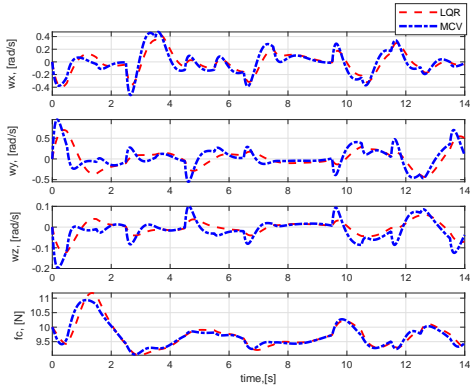}
    \caption{Comparison of input signals between the MCV and the LQR controllers for the straight line trajectory tracking. The MCV controller appears to have a faster response than the LQR controller.}
    \label{fig:input straght.}
\end{figure}

Simulation results for the straight line reference trajectory are illustrated through Figs~\ref{fig:traj_les}--\ref{fig:rmsles}, where we use $\gamma=0.75$. The trajectories of the LQR and the MCV controllers along with the nominal trajectory are plotted in Fig~\ref{fig:traj_les}, which demonstrate the effectiveness of MCV over LQR in reducing variance. We also conduct $50$ Monte Carlo simulations, where we incorporate different wind data and  calculate the variance and the RMS error at each reference point. The trajectory with  the MCV controller has smaller and smoother variance (see Fig~\ref{fig:varles}) and RMS error (see Fig~\ref{fig:rmsles}). A comparison of input signal $u$ is presented in Fig~\ref{fig:input straght.}. 

\begin{figure}
    \centering
    \includegraphics[width=7cm]{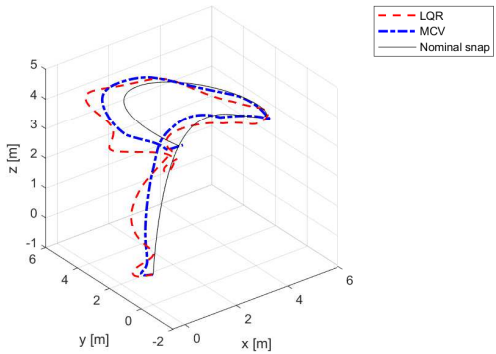}
    \caption{Comparison of circular trajectory tracking between the LQR and the MCV controllers (3D view).}
    \label{fig:3dlesz}
\end{figure}

\begin{figure}
    \centering
    \includegraphics[width=7cm]{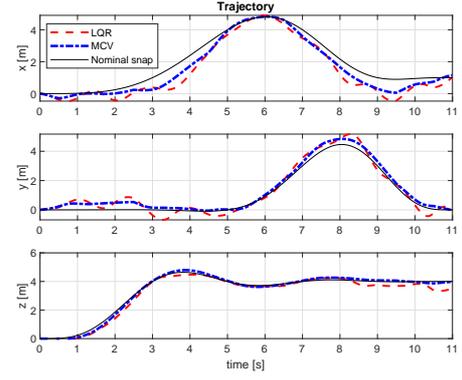}
    \caption{Comparison of circular trajectory tracking between the LQR and the MCV controllers for each direction.}
    \label{fig:cir_traj_lesz}
\end{figure}
\begin{figure}
    \centering
    \includegraphics[width=7cm]{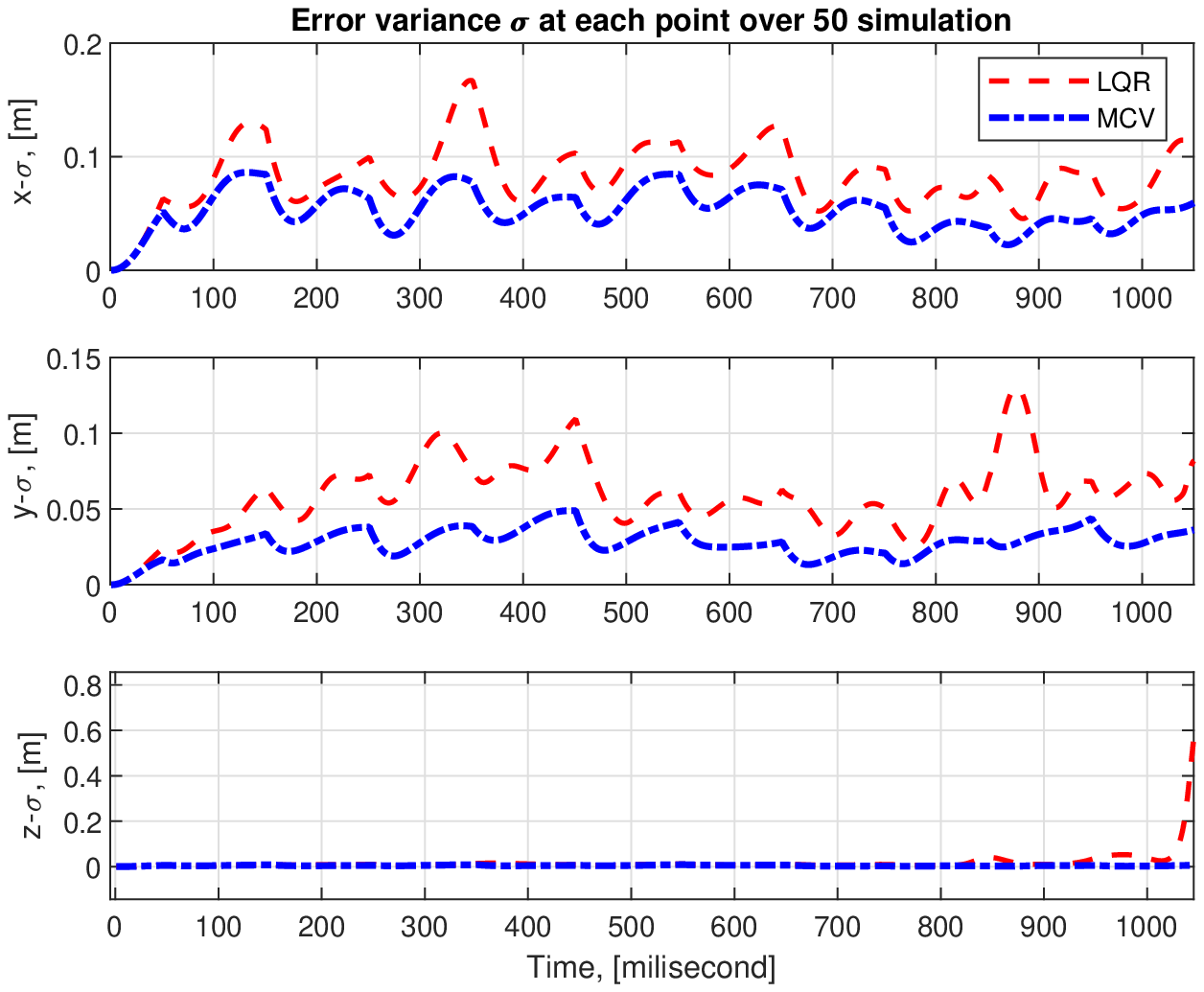}
    \caption{Comparison of error variance calculated at each point over 50 Monte Carlo simulations between the MCV and the LQR controllers for the circular trajectory tracking.}
    \label{fig:var_circ_lesz}
\end{figure}

\begin{figure}
    \begin{center}
    \includegraphics[width=7cm]{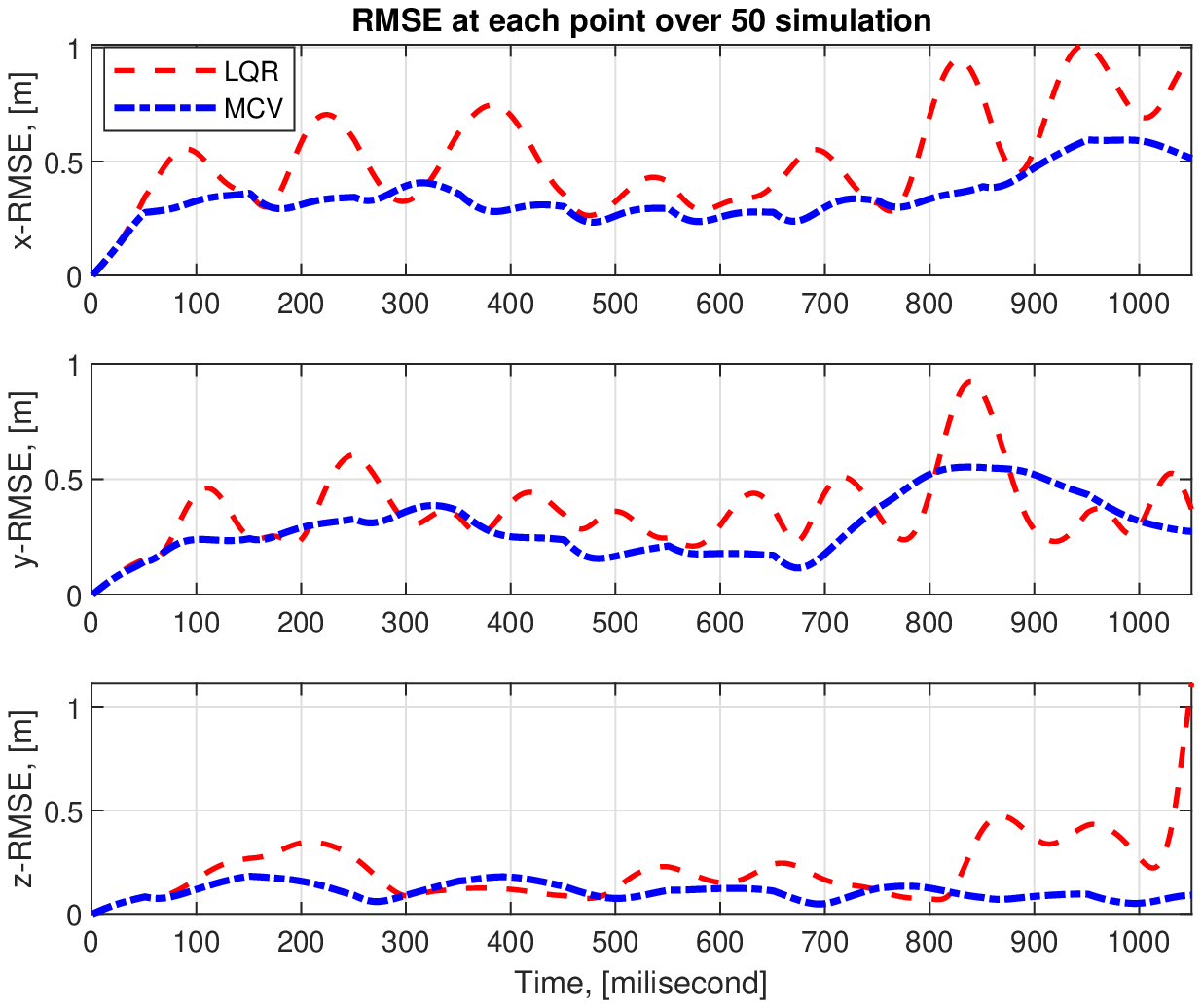}
    \caption{Comparison of the RMSE calculated at each point over 50 Monte Carlo simulations between the MCV and the LQR controllers for the circular trajectory tracking.}
    \label{fig:rms_circ_lesz}
\end{center}
\end{figure}

\begin{figure}
    \begin{center}
    \includegraphics[width=7cm]{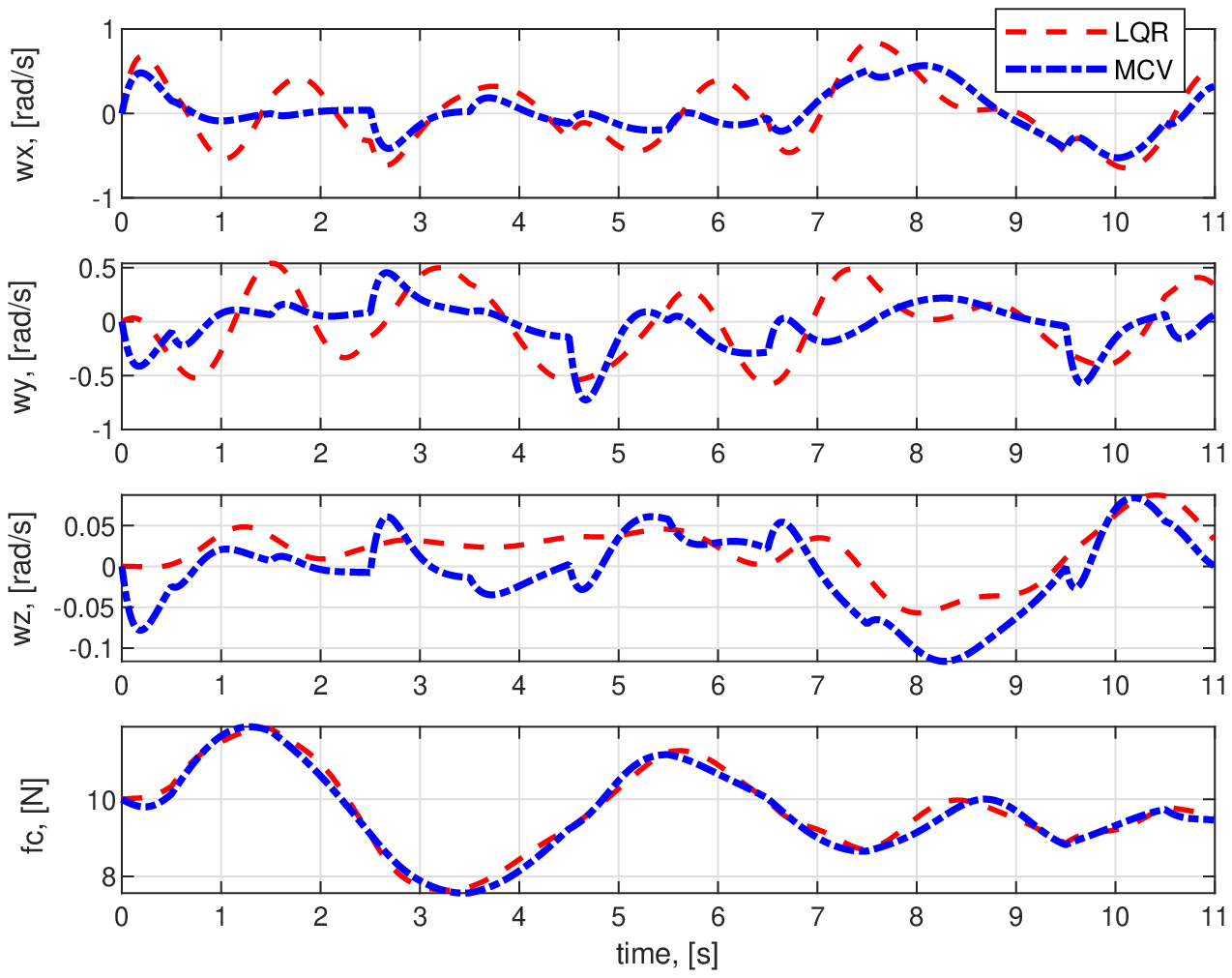}
    \caption{Comparison of input signals between the MCV and the LQR controllers for the circular trajectory tracking.}
    \label{fig:inp_circ}
\end{center}
\end{figure}

 We also generate a minimum snap circular trajectory as the reference trajectory. The trajectory passes through $(1,0,4),~(4,0,4),~(4,3,4)$ and $(1,3,4)$ starting from $(0,0,0)$. For the MCV controller, we set $\gamma=0.50$. The 3D circular tracking is provided in Fig~\ref{fig:3dlesz} and individual axis trajectory comparison in Fig~\ref{fig:cir_traj_lesz}. As expected,  the MCV controller results in lower variance and RMS error as illustrated in Figs \ref{fig:var_circ_lesz}--\ref{fig:rms_circ_lesz}. Overall we notice:
\begin{itemize}
    \item MCV reduces the variances as well as the RMS error of the trajectory. Although there still exists mean error, the variability is notably  reduced. %\HB{what do you mean by ``there are mean error the variability was notably  reduced"? I meant we could not reduces the mean error to zero but the variance was reduced to a better extent}https://www.overleaf.com/project/60733ecd42b721327d021de6
    \item In the straight line trajectory, the error variance in the $x$ direction is lower than $0.08$ m with the MCV where with the LQR the variance rises up to $0.245$ m, which is more than $3$ times than the MCV. In the $y$ direction, the MCV reduces the variance as much as $7$ times than the LQR (refer to Fig~\ref{fig:varles}).
   \item Although the circular trajectory exhibits higher variance than that of a straight line trajectory, the MCV controller still leads to  tracking with a smaller variance than LQR (refer to Fig~\ref{fig:cir_traj_lesz}). 
   \item From the input signal comparison in Fig~\ref{fig:input straght.} and Fig~\ref{fig:inp_circ}, we observe that in both cases the MCV controller responds earlier than the LQR controller. Also, the MCV controller input changes more as it reduces the noisy instances.

\end{itemize}

\section{Future Work}\label{future}
We design a Minimum Cost Variance controller for quadrotor control in a wind field. Our simulation results demonstrate its effectiveness in reducing the tracking error and variance in a turbulent wind field. We aim to implement the controller in higher-fidelity quadrotor simulator platforms, preferably in the ROS-Gazebo environment and simulate with spatial-temporal wind data. We are also exploring design methodologies to accommodate the nonlinearity in the dynamics into the controller.

\bibliography{root}             % bib file to produce

\begin{thebibliography}{24}
\providecommand{\natexlab}[1]{#1}
\providecommand{\url}[1]{\texttt{#1}}
\providecommand{\urlprefix}{URL }
\expandafter\ifx\csname urlstyle\endcsname\relax
  \providecommand{\doi}[1]{doi:\discretionary{}{}{}#1}\else
  \providecommand{\doi}{doi:\discretionary{}{}{}\begingroup
  \urlstyle{rm}\Url}\fi

\bibitem[{Allison et~al.(2020)Allison, Bai, and Jayaraman}]{allison2020wind}
Allison, S., Bai, H., and Jayaraman, B. (2020).
\newblock Wind estimation using quadcopter motion: A machine learning approach.
\newblock \emph{Aerospace Science and Technology}, 98, 105699.

\bibitem[{Beare et~al.(2006)Beare, Macvean, Holtslag, Cuxart, Esau, Golaz,
  Jimenez, Khairoutdinov, Kosovic, Lewellen et~al.}]{beare2006intercomparison}
Beare, R.J., Macvean, M.K., Holtslag, A.A., Cuxart, J., Esau, I., Golaz, J.C.,
  Jimenez, M.A., Khairoutdinov, M., Kosovic, B., Lewellen, D., et~al. (2006).
\newblock An intercomparison of large-eddy simulations of the stable boundary
  layer.
\newblock \emph{Boundary-Layer Meteorology}, 118(2), 247--272.

\bibitem[{Bisheban and Lee(2018)}]{bisheban2018geometric}
Bisheban, M. and Lee, T. (2018).
\newblock Geometric adaptive control for a quadrotor uav with wind disturbance
  rejection.
\newblock In \emph{2018 IEEE Conference on Decision and Control (CDC)},
  2816--2821. IEEE.

\bibitem[{Bryan and Fritsch(2002)}]{bryan2002benchmark}
Bryan, G.H. and Fritsch, J.M. (2002).
\newblock A benchmark simulation for moist nonhydrostatic numerical models.
\newblock \emph{Monthly Weather Review}, 130(12), 2917--2928.

\bibitem[{Computational and Laboratory(2017)}]{computational2017cheyenne}
Computational and Laboratory, I.S. (2017).
\newblock Cheyenne: Hpe/sgi ice xa system (university community computing).

\bibitem[{Davoudi et~al.(2020)Davoudi, Taheri, Duraisamy, Jayaraman, and
  Kolmanovsky}]{davoudi2020quad}
Davoudi, B., Taheri, E., Duraisamy, K., Jayaraman, B., and Kolmanovsky, I.
  (2020).
\newblock Quad-rotor flight simulation in realistic atmospheric conditions.
\newblock \emph{AIAA Journal}, 58(5), 1992--2004.

\bibitem[{Deardorff(1980)}]{deardorff1980stratocumulus}
Deardorff, J.W. (1980).
\newblock Stratocumulus-capped mixed layers derived from a three-dimensional
  model.
\newblock \emph{Boundary-Layer Meteorology}, 18(4), 495--527.

\bibitem[{Ding and Wang(2018)}]{ding2018robust}
Ding, L. and Wang, Z. (2018).
\newblock A robust control for an aerial robot quadrotor under wind gusts.
\newblock \emph{Journal of Robotics}, 2018.

\bibitem[{Foehn and Scaramuzza(2018)}]{foehn2018onboard}
Foehn, P. and Scaramuzza, D. (2018).
\newblock Onboard state dependent lqr for agile quadrotors.
\newblock In \emph{2018 IEEE International Conference on Robotics and
  Automation (ICRA)}, 6566--6572. IEEE.

\bibitem[{Gill and D'Andrea(2017)}]{gill2017propeller}
Gill, R. and D'Andrea, R. (2017).
\newblock Propeller thrust and drag in forward flight.
\newblock In \emph{2017 IEEE Conference on Control Technology and Applications
  (CCTA)}, 73--79. IEEE.

\bibitem[{Hamadi et~al.(2019)Hamadi, Lussier, Fantoni, Francis, and
  Shraim}]{hamadi2019observer}
Hamadi, H., Lussier, B., Fantoni, I., Francis, C., and Shraim, H. (2019).
\newblock Observer-based super twisting controller robust to wind perturbation
  for multirotor uav.
\newblock In \emph{2019 International Conference on Unmanned Aircraft Systems
  (ICUAS)}, 397--405. IEEE.

\bibitem[{Jiang and Shu(1996)}]{jiang1996efficient}
Jiang, G.S. and Shu, C.W. (1996).
\newblock Efficient implementation of weighted eno schemes.
\newblock \emph{Journal of computational physics}, 126(1), 202--228.

\bibitem[{{Mellinger} and {Kumar}(2011)}]{5980409}
{Mellinger}, D. and {Kumar}, V. (2011).
\newblock Minimum snap trajectory generation and control for quadrotors.
\newblock In \emph{2011 IEEE International Conference on Robotics and
  Automation}, 2520--2525.
\newblock \doi{10.1109/ICRA.2011.5980409}.

\bibitem[{Sain(1965)}]{sain1965minimal}
Sain, M.K. (1965).
\newblock On minimal-variance control of linear systems with quadratic loss.
\newblock Technical report, ILLINOIS UNIV URBANA COORDINATED SCIENCE LAB.

\bibitem[{Sain et~al.(1995)Sain, Won, and Spencer}]{sain1995cumulants}
Sain, M.K., Won, C.H., and Spencer, B. (1995).
\newblock Cumulants in risk-sensitive control: The full-state-feedback cost
  variance case.
\newblock In \emph{Proceedings of 1995 34th IEEE Conference on Decision and
  Control}, volume~2, 1036--1041. IEEE.

\bibitem[{Sierra and Santos(2019)}]{sierra2019wind}
Sierra, J.E. and Santos, M. (2019).
\newblock Wind and payload disturbance rejection control based on adaptive
  neural estimators: application on quadrotors.
\newblock \emph{Complexity}, 2019.

\bibitem[{Smagorinsky(1963)}]{smagorinsky1963general}
Smagorinsky, J. (1963).
\newblock General circulation experiments with the primitive equations: I. the
  basic experiment.
\newblock \emph{Monthly weather review}, 91(3), 99--164.

\bibitem[{Tran et~al.(2015)Tran, Bulka, and Nahon}]{tran2015quadrotor}
Tran, N.K., Bulka, E., and Nahon, M. (2015).
\newblock Quadrotor control in a wind field.
\newblock In \emph{2015 International Conference on Unmanned Aircraft Systems
  (ICUAS)}, 320--328. IEEE.

\bibitem[{Tran et~al.(2021)Tran, Santoso, and Garratt}]{tran2021adaptive}
Tran, V.P., Santoso, F., and Garratt, M.A. (2021).
\newblock Adaptive trajectory tracking for quadrotor systems in unknown wind
  environments using particle swarm optimization-based strictly negative
  imaginary controllers.
\newblock \emph{IEEE Transactions on Aerospace and Electronic Systems}.

\bibitem[{Von~Karman(1948)}]{von1948progress}
Von~Karman, T. (1948).
\newblock Progress in the statistical theory of turbulence.
\newblock \emph{Proceedings of the National Academy of Sciences of the United
  States of America}, 34(11), 530.

\bibitem[{Wang et~al.(2016)Wang, Song, Huang, and Tang}]{wang2016trajectory}
Wang, C., Song, B., Huang, P., and Tang, C. (2016).
\newblock Trajectory tracking control for quadrotor robot subject to payload
  variation and wind gust disturbance.
\newblock \emph{Journal of Intelligent \& Robotic Systems}, 83(2), 315--333.

\bibitem[{Won et~al.(2003)Won, Sain, and Liberty}]{won2003infinite}
Won, C.H., Sain, M.K., and Liberty, S.R. (2003).
\newblock Infinite-time minimal cost variance control and coupled algebraic
  riccati equations.
\newblock In \emph{Proceedings of the 2003 American Control Conference, 2003.},
  volume~6, 5155--5160. IEEE.

\bibitem[{Yang et~al.(2017)Yang, Cheng, Xia, and Yuan}]{yang2017active}
Yang, H., Cheng, L., Xia, Y., and Yuan, Y. (2017).
\newblock Active disturbance rejection attitude control for a dual closed-loop
  quadrotor under gust wind.
\newblock \emph{IEEE Transactions on control systems technology}, 26(4),
  1400--1405.

\bibitem[{Zhang et~al.(2016)Zhang, Zhou, Zhao, Dai, and Zhou}]{zhang2016three}
Zhang, C., Zhou, X., Zhao, H., Dai, A., and Zhou, H. (2016).
\newblock Three-dimensional fuzzy control of mini quadrotor uav trajectory
  tracking under impact of wind disturbance.
\newblock In \emph{2016 International Conference on Advanced Mechatronic
  Systems (ICAMechS)}, 372--377. IEEE.

\end{thebibliography}
\end{document}